\documentclass[prl,final,twocolumn,showpacs]{revtex4}

\usepackage{graphicx}
\usepackage{dcolumn}

\begin{document}

\title{Evidence of spontaneous spin polarized transport in magnetic nanowires}

\author{Varlei Rodrigues}
\author{Jefferson Bettini}
\author{Paulo C. Silva}
\author{Daniel Ugarte}
\email{ugarte@lnls.br}
\affiliation{Laborat\'{o}rio Nacional de Luz S\'{\i}ncrotron, C.P. 6192, 13084-971 Campinas SP, Brazil}
\date{\today}

\begin{abstract}
The exploitation of the spin in charge-based systems is opening revolutionary opportunities
for device architecture. Surprisingly, room temperature electrical transport through 
magnetic nanowires is still an unresolved issue. Here, we show that ferromagnetic (Co) 
suspended atom chains spontaneously display an electron transport of half a conductance 
quantum, as expected for a fully polarized conduction channel. Similar behavior has 
been observed for Pd (a quasi-magnetic 4d metal) and Pt (a non-magnetic 5d metal). 
These results suggest that the nanowire low dimensionality reinforces or induces 
magnetic behavior, lifting off spin degeneracy even at room temperature and zero 
external magnetic field. 
\end{abstract}


\maketitle

It is expected that the new generation of devices will exploit spin dependent effects, 
what has been called "spintronics" \cite{1}. In this sense, the role of low dimensionality 
in the magnetic properties of materials will become a fundamental issue for combining 
the standard miniaturization of microelectronics and spin phenomena. From a practical 
point of view, spintronic devices must exploit different quantum properties without the 
need of cryogenic temperatures. This fact immediately renders metal nanowires (NW's) a 
very attractive system because they show quantum conductance effects at room 
temperature \cite{2}. From a experimental point of view, NW's can be easily generated by 
putting in contact two metal surfaces, which are subsequently pulled apart; during 
the elongation, the conductance ($G$) displays flat plateaus and abrupt jumps of approximately 
a conductance quantum $G_{0} = 2 e^{2}/h$, where $e$ is the electron charge and $h$ is Planck's 
constant \cite{2}; the factor 2 is due to the spin degeneracy. Recently, this kind of 
studies has lead to the discovery that the ultimate NW's show a structure of suspended 
linear chain of atoms (LCA) \cite{3,4,5}, whose conductance is equal to 1 $G_{0}$ 
for monovalent 
metals such as Au \cite{3,4,6} or Ag \cite{7}. On the other side, magnetic materials have not 
yet been studied in detail and the possible lift of spin degeneracy in 
magnetic NW's represents still an open question \cite{2,8}.

Here, we have analyzed the room temperature electronic transport properties of atom-size metallic 
wires made of magnetic and non-magnetic metals using an ultra-high-vacuum (UHV, pressure $\leq$ 10$^{-8}$ Pa) 
mechanically controllable break junction system (MCBJ) \cite{2, 6, 7}, which is a well 
established technique to study the conductance of nanostructures. 
In this approach, a macroscopic wire is glued in a flexible substrate in two points; then it is rendered fragile between the two fixing parts by an incomplete cut. By bending the substrate \textit{in situ} in UHV, we break the wire and produce two clean metal surfaces. Using the same bending movement, the fresh tips are put together and separated repeatedly to generate and elongate NW's. It must be emphasized that the extreme cleanness of the environment and the sample itself are essential to get reliable and reproducible conductance data from atomic-size contacts \cite{6, 7}. 

In a MCBJ, the electrical transport of the metal NW's is measured using a two-point configuration; this implies that the conductance measurement probes the NW itself (the narrowest region of the contact) and the two leads (apexes). Although, the NW should show conductance quantization ($G = \alpha   G_{0}$, where $\alpha$ is an integer), the electron reservoir-apex-NW coupling may act as an additional serial resistance, diminishing the conductance. Experimentally, $\alpha$ is frequently close to integer values but slightly lower. 

In these experiments, an additional difficulty arises from the fact that a variety of conductance evolutions are observed. In fact, the structure of the NW or relative crystallographic orientation of the apexes cannot be controlled; then, each conductance measurement corresponds to a new NW with a different atomic arrangement. To overcome this difficulty, most NW studies rely on the analysis of average behaviors of many conductance curves. The most frequently used procedure consists in building histograms from each individual electrical transport measurement, where occurrence of each conductance value is plotted (in this way a conductance plateau becomes an histogram peak). Subsequently, the so-called global histogram is constructed by the linear addition of individual histograms from a series of measurements. The presence of peaks close to integer multiples of the conductance quantum has been considered as the proof of quantized conductance in metal NW's \cite{2}.

The atomic structure of NW's generated by mechanical 
stretching has been studied using independent experiments based on time-resolved 
high resolution transmission electron microscopy (HRTEM), where the NW's were 
generated and elongated inside the microscope following the method reported 
by Kondo and Takanayagi \cite{9}. NW's are generated by the following procedure: the microscope electron beam is increased to a current density of $\sim$ 120 A/cm2 and focussed on a self-supported metal film to perforate and grow neighboring holes. When two holes are very close, a nanometric bridge is formed between them. When these bridges are very thin (1-2 nm) and close to rupture, the electron beam intensity is reduced to its conventional value (10-30 A/cm2) in order to perform the real-time imaging with atomic resolution. When the beam current is too high (as during the hole generation step), the film vibrates and no atomic imaging can be performed. The procedure described above has allowed us to generate NW's with a remarkable stability. In fact, the NW, its apexes and the surrounding regions are all parts of a unique metal film and form a monolithic block. NW's formed by a few atomic layers usually show a long lifetime in the range of minutes. Although this stability, the NW's elongate spontaneously, get thinner and, then break due to the relative slow movement of the NW apexes. This apex displacement is probably due to a film deformation induced by thermal gradients or just by low frequency vibration of the thin metallic film membrane, as usually observed in TEM thin film work \cite{6,7}. 

The metal films (thickness 15 nm for Co, 5-6 nm for Pd and Pt) have been obtained by thermal or electron beam evaporation of pure metal on a substrate (usually NaCl crystals or freshly cleaved mica). Subsequently, the films are detached from the substrate by floating them in water; next, the sample is collected on a TEM holey carbon grid, remaining self-supported on the regions hanging over the holes. All images were acquired close to Scherzer defocus \cite{10} using a high sensitivity TV camera (Gatan 622SC, 30 frame/s) associated with a conventional video recorder. The images were obtained by digitizing the video film \textit{a posteriori}; in order to enhance signal-to-noise ratio, several frames (3-5) are usually added. This procedure has shown to be very efficient and, has enabled us to register the NW real-time formation, evolution and rupture, even for low atomic number metals as for example Co. 

The steps mentioned above constitute the basic procedure for \textit{in situ} NW observation. However, we must emphasize that the study of NW's has demonstrated to be a new challenge for each metal. Each material has required a particular setting, from the sample preparation (thin film thickness, etc.) to the NW generation and imaging. For this kind of experiments, noble metals seem to be a well-behaved case; on the other side, metals such as Co are more complicated, mainly because they are very reactive. We have prevented sample oxidation by evaporating sequentially carbon-metal-carbon layers on a substrate. This "metal sandwich" is detached from the substrate and collected on the TEM grid, as described above. Inside the TEM, the carbon layers are removed from a sample region using an intense electron irradiation, what can be very time consuming (6-8 hs.). 

Figure 1 shows typical behaviors of the conductance during elongation and thinning 
of Co NW's; also a histogram of occurrence of each conductance value  
(global histogram \cite{2}) during a series of measurements is displayed in order to 
derive a statistical analysis of NW behavior. A quick analysis shows a major 
difference with previously studied metals 
(Au \cite{2, 6}, Ag \cite{2, 7}, Cu \cite{2}, Na \cite{11}, etc.): 
the last conductance plateau  
(or the lowest conductance peak of the global histogram) attributed to the 
thinnest Co wire is observed at $\sim$  0.5 $G_0$ instead of $\sim$  1 $G_0$. 

\begin{figure}
\includegraphics[width = 6 cm]{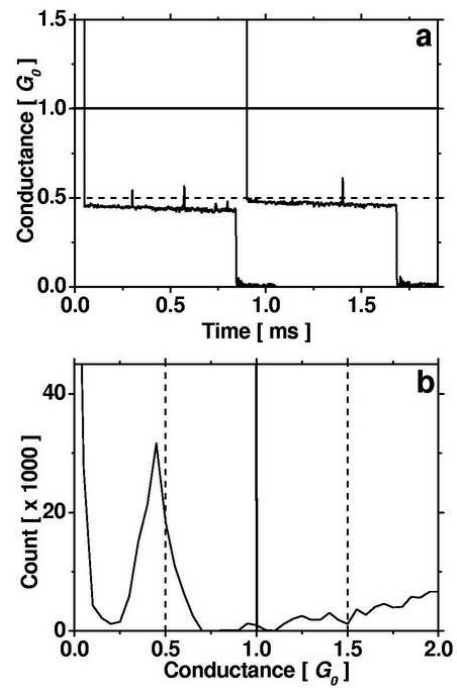}
\caption{
UHV-MCBJ conductance measurements of Co NW's at room temperature and without external 
magnetic field. (a) Typical electrical transport curves showing conductance plateaus. 
(b) Global histogram exhibiting the statistical conductance behaviour of a sequence 
of NW generations. Note that the plateaus in (a) and the lowest peak in (b) are 
located at $\sim$ 0.5 $G_0$.
}
\end{figure}

In order to understand the origin of the 0.5 $G_0$ conductance, it is essential to 
determine the atomic arrangement of the thinnest possible Co NW. HRTEM imaging 
has shown that just before rupture, Co wires adopt a LCA configuration 
(see Fig. 2a), which must be associated to the last conductance plateau 
at 0.5 $G_0$. It must be noted that the global histogram, in Fig. 1b, does 
not display a 1 $G_0$ peak, as could be expected by the addition of 0.5 $G_0$ steps. 
However, this is not surprising because it is already well understood that 
conductance curves are a signature of the structural evolution during the NW 
stretching \cite{6}. Then, some conductance values may not be observed in a transport 
experiment, if there is not a stable atomic structure sustaining that particular 
conductance value \cite{7, 12}.

\begin{figure}
\includegraphics[width = 6 cm]{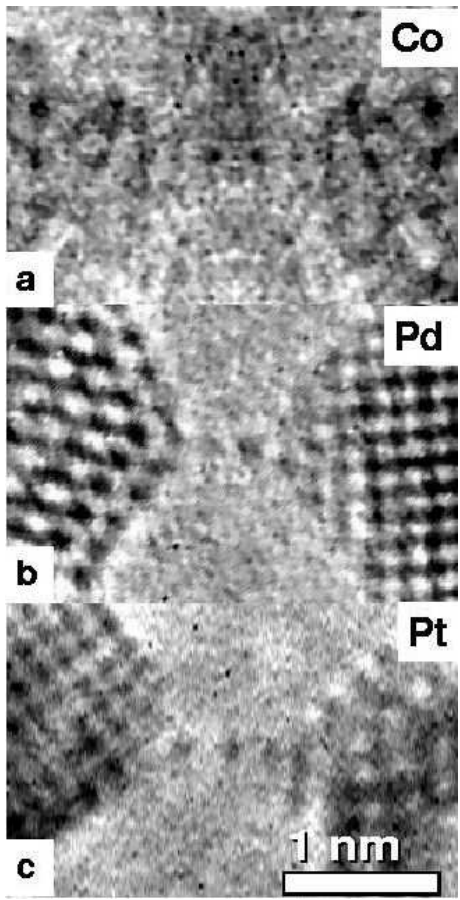}
\caption{
HRTEM atomic resolved images showing the formation of suspended chains of atoms 
just before the contact rupture. (a) Co. (b) Pd. (c) Pt.
}
\end{figure}

Conductance plateaus and jumps of 0.5 $G_0$ are observed for constrictions generated in 
bi-dimensional electron gazes when a strong magnetic field is used to lift the 
spin degeneracy; in fact, features at 0.5 $G_0$ are considered a signature of a 
spin polarized current generation \cite{1, 2, 13, 14, 15}.  Our results enable us to deduce 
that this phenomenon occurs spontaneously for one-atom-thick ferromagnetic wires 
in zero magnetic field and at room temperature. The polarized current can be 
originated either from a wide spin gap at the  narrowest wire region, or from 
the fact that the apexes contacting the atom chain have identical magnetic 
orientation (in other terms, magnetic domain walls may be expelled from the nanometric 
constriction). Further studies involving temperature or magnetic field dependence are in progress to elucidate this point. 
This behavior was not only observed in Co wires, but also in NW's made of 
other ferromagnetic 3d transition metals such as Fe and Ni (not shown here). 
In contrast, it was not verified in similar experiments on Cu, a very close 
non-magnetic 3d transition metal \cite{2, 16}. This kind of analysis 
can only be applied to the last conductance plateau (0.5 $G_0$) and, we are not able 
to derive further insights on the existence of polarized current in thicker NW's 
(or higher conductance), because the superposition of possible spin 
polarized and unpolarized conduction channels cannot be discriminated from the 
conductance value. 

\begin{figure}
\includegraphics[width = 6 cm]{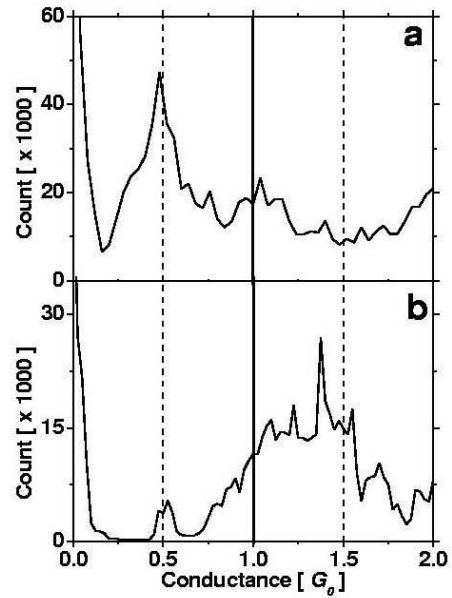}
\caption{
Global histograms of conductance for (a) Pd and (b) Pt measured at room temperature 
and without external magnetic field. Note that the peak corresponding the lowest 
conductance value is located at $\sim$ 0.5 $G_0$ for both metals.
}
\end{figure} 

One-atom-thick wires represent the smallest achievable 1D system; then, they should 
be expected to show new and unexpected physical behaviors. For example, magnetic 
ordering may occur in low-dimensional systems of nonmagnetic materials \cite{17}. It has been theoretically predicted that infinite 1D spin systems would not display magnetism \cite{18}, 
nevertheless Gambardella \textit{et al} \cite{19} have recently revealed magnetic ordering in chains of Co atoms deposed on Pt and spin blocks of 15 atoms at 45 K were reported. Certainly in this case, the Pt substrate must play an non-negligible role, but these results suggest that ferromagnetism could be expected in the short suspended chains analyzed in this work. On the other hand, it is surprising that the magnetic ordering seem to occur even at room temperature.
In this 
sense, it will be very interesting to check if LCA's of quasi-magnetic metals become 
magnetic and, may show spin polarized conductance channels. The obvious test case 
would be the 4d transition metal Pd ([Kr] 4d$^{10}$), where 2-D thin films and 0-D clusters 
have already been reported to become magnetic \cite{20,21}. Firstly, we have used HRTEM to 
show that the thinnest Pd nanowire displays a LCA structure (see Fig. 2b). 
Subsequently, UHV-MCBJ experiments revealed that the global 
histogram shows the lowest conductance peak (associated to the LCA structure) 
at 0.5 $G_0$ (see Fig. 3a), as expected when a spin polarized current is allowed to 
travel through the Pd LCA's. 

An extension of the previous studies to the 5d transition metals leads to Pt as 
a test metal candidate. For that row, Pt shows a stronger localization of the d 
wave function, although 5d metals have a very small exchange integral \cite{22}. 
Concerning atomic structure, Pt NW's have already been shown to form LCA's \cite{2, 23}, 
see example in Fig 2c. The results of conductance measurements of Pt NW's are 
shown in Fig. 3b, where a well-defined peak at 0.5 $G_0$ can be easily recognized 
in the global histogram. The formation of a spin polarized current in Pt atom 
chains may seem surprising because the possible occurrence of magnetism in 5d 
transition metals has usually been neglected \cite{22}. However, molecular beam 
photo-detachment studies of very small (2-3 atoms) Pt clusters have shown an 
anomalous behavior, somewhat following the tendency observed for magnetic 
clusters (Ni or Pd) \cite{20}. Also, we must keep in mind that in this size regime 
metal clusters adopt a linear structure \cite{24}, very close to the atomic 
arrangement in LCA's. In addition, recent \textit{ab-initio} calculations have predicted 
the generation of a magnetic state in highly elongated linear Pt chains
\cite{25,26}; the need of significant bond elongation could explain why the 
0.5 $G_0$ peak is rather small for Pt, while dominant for Co and Pd.

In summary, we have shown that suspended chains of atoms made of ferromagnetic 
3d transition metals display a conductance compatible with a fully polarized 
conduction channel (0.5 $G_0$) at room temperature and without the need of external 
magnetic field. The 1D nature of LCA's also induces a similar conductance behavior 
in suspended chains of the quasi-magnetic 4d metal Pd and, a non-magnetic 5d metal 
as Pt. These results open a wealth of new opportunities to get a deeper 
understanding of spin dynamics or spin control in nanostructures and, will 
have important implications for the development of future spintronic devices. 
Although, the practical application of linear chain of atoms will be rather 
hard to achieve, their atomic structure provides a very good insight for designing 
and synthesizing new organometallic molecules containing magnetic atoms for 
applications in molecular electronics exploiting spin effects.

We thank M. Knobel for stimulating discussions. This work was supported LNLS, CNPq and FAPESP.

\end{document}